\documentclass{JHEP} \usepackage{epsf} 

\newcommand{\beq}{\begin{equation}}
\newcommand{\beql}[1]{\begin{equation}\label{eq:#1}}
\newcommand{\eeq}{\end{equation}} 
\newcommand{\beqn}{\begin{eqnarray}}
\newcommand{\eeqn}{\end{eqnarray}} 
\newcommand{\eq}[1]{(\ref{eq:#1})}
\newcommand{\pa}{\partial}
\newcommand{\tr}{{\rm tr}}
\newcommand{\comment}[1]{}

\newcommand{\sigmod}{$\sigma$-model}
\newcommand{\phf}{{\rm P}_{1/2}}
\newcommand{\bphf}{\bar {\rm P}_{1/2}}
\newcommand{\CO}{{\cal O}}

\newcommand{\calD}{{\cal D}}
\newcommand{\pasl}{\pa\kern-.55em /}
\newcommand{\Dsl}{D\kern-.65em /}
\newcommand{\ksl}{k\kern-.5em /}
\newcommand{\sX}{{\bf X}}
\newcommand{\CS}{{\cal S}}
\newcommand{\iden}[1]{{\cal I}^{\dot {#1}}}
\newcommand{\tiden}[1]{\tilde {\cal I}^{\dot {#1}}}

\newcommand{\littlefig}[2]{
	\epsfxsize=#2in
	\epsfbox{#1}
}

\title{Quantization of Superstrings in Ramond-Ramond Backgrounds}
\author{David Berenstein\thanks{\email{berenste@hepux0.hep.uiuc.edu}}
 and Robert G. Leigh\thanks{\email{rgleigh@uiuc.edu}} \\ Department
of Physics\\ University of Illinois at Urbana-Champaign\\
 Urbana, IL 61801 }
\abstract{
We present a perturbative study of Ramond-Ramond backgrounds in the NSR
formalism.  We show how to perform \sigmod\ computations
and discuss in detail the structure of the BRST charge and picture 
changing operators. Contact terms
play a vital role in the analysis. We also give evidence for a two
loop non-renormalization theorem for the background beta functions.
}
\preprint{ILL-(TH)-99-05\\ hep-th/9910145}
\begin{document}

\section{Introduction}

The quantization of superstrings in Ramond-Ramond ($RR$) backgrounds is an
important theoretical problem, as it has direct significance to the
large $N$ behaviour of strongly coupled gauge theories \cite{Juan}. For
a review see \cite{AGMOO}. Several attacks at this problem have been
made in the literature. The Green-Schwarz formalism for the interesting
case of $AdS_5\times S_5$ has been studied in the classical theory by a
number of authors,\cite{MT,KRo} and aspects of the quantization of this
action as an expansion around flat spacetime has been considered in Ref.
\cite{ARMR}. Light cone gauge has been considered further in Ref.
\cite{M}. The $RR$ background with $AdS_3\times S_3$ geometry has been
considered in Ref. \cite{BVW,Berk} and from a Green-Schwarz perspective
in Refs. \cite{Pes,Rahm}.

Either way one looks at the quantization problem, there are technical
questions which need to be answered in order to have a string
quantization. In a Green-Schwarz formulation of the superstring, we only
know how to perform a quantization in the lightcone gauge. For a general
background, one can not expect to have a well-defined notion of
lightcone. Moreover, if there is a choice of lightcone, it is not clear
that in a curved background one can decouple the lightcone degrees of
freedom from the rest of the worldsheet CFT.

If one attempts to work in a covariant formulation of the superstring,
one must confront technical questions about the worldsheet CFT. In
particular we are used to dealing with $NS$ backgrounds, which respect
worldsheet supersymmetry. The BRST structure and the picture changing
operations follow directly from worldsheet supersymmetry, so one can
calculate the constraints on physical states without difficulty. As we
discuss below, an $RR$ background breaks worldsheet supersymmetry.
Moreover, the usual picture-changing operator is defined in terms of the
worldsheet supersymmetry generator; in $RR$ backgrounds, one must find
an alternate way to remove the ambiguities of picture. It is important
to stress here that although we are used to thinking of the superstring
in flat spacetime as a simple theory, it is {\it not} a free theory.
Rather, it is usually described via a free field realization; the price
we pay for that is the complications of screening and picture. It is
then not surprising that we must carefully consider the modifications to
these structures in $RR$ backgrounds.

It is the purpose of this paper to attack these problems and suggest an
answer as to how the BRST structure and the picture changing operations
are modified by the $RR$ background within conformal perturbation
theory.

In an earlier paper,\cite{first} we began a perturbative analysis of
superstring theory in Ramond-Ramond backgrounds using the familiar
Ramond-Neveu-Schwarz ($RNS$) formalism. We consider (Type IIB) string
theory on an arbitrary smooth target space with non-zero Ramond-Ramond
fieldstrength background. For simplicity, we assume the dilaton is
constant,\footnote{The significance of this statement in the context of
the \sigmod\ will be pointed out later. The spacetime significance is
obvious.} and all other moduli are set to zero. We also assume that the
background is such that the $\alpha'$ expansion and the string loop
expansion is well-behaved. The closed string \sigmod\ was derived
explicitly through a process of summing over open string worldsheets in
a D-brane\cite{LP} background. The \sigmod\ action is deduced by
requiring conformal invariance order by order in the expansion in $1/r$,
the distance from the branes, an application of the Fischler-Susskind
mechanism.\cite{WFLS} This \sigmod\ has the form
\beql{sigmodaction}
S=\int d^2zd^2\theta\ g_{\mu\nu}({\bf X}) D{\bf X}^\mu\bar D{\bf X}^\nu
+\int d^2 z\ \phf\bphf h_{\alpha\beta}(X) 
 e^{-\phi/2}S^\alpha e^{-\bar\phi/2}\tilde S^\beta
\eeq
In this expression, the fields $\phi$ and $\bar\phi$ are those found in
the bosonization of the $\beta\gamma$ superghost system of the flat
spacetime theory, and $\alpha,\beta$ are spacetime chiral spinor indices. 
Further notational details may be found in the Appendix. The formal
dimension zero operators $\phf$ and $\bphf$ are thought of as the
(anti-)holomorphic square root of the picture-changing operator. These
remove the ambiguity of ghost-picture, so that to each order in
perturbation theory the $RR$-background vertex does not affect the
pictures of external states. For the low order calculations presented in
Ref. \cite{first}, the lack of knowledge of the detailed form of $\phf$
was not important, as for any (non-zero) S-matrix element, these
operators always appear in pairs. In the present paper, we confront this
issue more directly and discuss how to consistently deal with these
operators, order by order in perturbation theory.

An important aspect of these backgrounds is that the CFT in no sense
splits holomorphically. Aspects of this were pointed out in
\cite{first}, such as the realization of spacetime symmetry currents and
their algebras. Regardless, we will show in this paper that conformal
invariance may be consistently maintained, and a holomorphic stress
tensor constructed order by order. Furthermore, the BRST operator may be
constructed; the resulting restrictions on physical states are
consistent with spacetime supersymmetry.

Much of the discussions presented in this paper involve a careful analysis
of contact terms. Indeed, the \sigmod\ action should be supplemented with
a consistent choice of contact terms in order to define the theory.
(In fact, as we will discuss, the presentation of the $NSNS$ part of
the action as an integral over superspace corresponds to a choice of
contact terms, valid at lowest order.) The guiding principles for 
choosing contact terms will be worldsheet conformal invariance and
spacetime supersymmetry \cite{GS}. 

In the presence of the $RR$ background, we expect operator mixing
between operators of $NS$ and $R$ type. Thus it is a possibility that
the BRST charge and picture-changing operator itself contains $RR$
contributions. In the latter case, such effects would seriously
complicate the perturbative expansion.

One of the principal results of this paper is that the corrections to
the picture-changing operator are of a very simple form at low
orders--it is simply covariantized  with respect to the background
metric. In particular, there appears to be no $RR$ contribution at tree
level, and the ghost-picture structure of this theory is then very much
as it is in flat spacetime.  We will comment more fully on this subject
in later sections.

In general, because of the conformal invariance of the background, we
expect that there exists a nilpotent BRST current which gets
contributions from both $NS$ and $R$ sectors. The results of this paper
are consistent with the assumption that only the $NS$ part of $J_{BRST}$
contributes to $P_{+1}$ at tree level. In particular, the
$\eta$-dependence of $Q_{BRST}$ is severely restricted.
\begin{equation}
J_{BRST} = \gamma g_{\mu\nu}(X)\psi^\mu\pa X^\nu +\hbox{ghosts} +\ldots
= e^\phi \eta\ G_{matter} +\ldots
\end{equation}
where the pure ghost pieces are independent of the background.  

Since the \sigmod\ is interacting, all statements are perturbative in
$\alpha'$; to verify these equations we use an expansion around a well
known background (we choose flat backgrounds for our example), which is
devoid of the $RR$ field strength and which satisfies the string
equations of motion.

The paper is organized as follows. In the next section, we discuss the
issue of (the absence of) worldsheet supersymmetry in $RR$ backgrounds.
In Section \ref{sec:contact}, we discuss the importance of contact terms
in string perturbation theory. In Section \ref{sec:oneloop}, we present
the details of one-loop \sigmod\ computations. In Section \ref{sec:brst},
we discuss the form of the $BRST$ charge. After a short review of $BRST$
symmetry within the standard superstring, we show explicitly that the
$RR$ background does not modify the picture-changing operator at tree
level, thus validating the one-loop computations performed in
Section \ref{sec:oneloop}. We argue however that at one loop there are 
non-trivial contact terms
which imply that the $BRST$ charge is modified by the $RR$ background.
This follows from the $BRST$ invariance of the background.
We find a description in which the the BRST charge and the stress tensor
are modified holomorphically. Using this description, we are able to
show that the resulting $BRST$ constraints on spacetime boson and fermion
states are precisely equivalent to the respective equations of motion. 
Finally, in Section \ref{sec:twoloop}, we consider briefly higher loops
in \sigmod\ perturbation theory. We give worldsheet evidence that two
and three loop contributions to the $\beta$-functions vanish. This 
result is of course consistent with expectations from spacetime
arguments. We conclude the paper with additional comments.

\section{Worldsheet Supersymmetry}\label{sec:wssusy}

In eq. \eq{sigmodaction}, we have written the $NSNS$ term in terms of
worldsheet superspace for brevity. It is important to consider here the
significance, if any, of worldsheet supersymmetry. To this end, recall
the sum over spin structures of the flat spacetime superstring.
\beql{spinstruct}
Z=\sum_\sigma Z_\sigma
\eeq
Superconformal symmetry acts simply within each spin structure
$Z_\sigma$, but not of course in any sense on the whole. Note that the
superconformal transformations have branch cuts at the positions of $RR$
insertions (equivalently, the OPE is non-local).

In summing over worldsheets in a D-brane background, we are effectively
summing over insertions of tadpoles produced by the D-brane. As is
well-known,\cite{P} there are tadpoles for both $NSNS$ fields as well
as $RR$ fields. Thus summation over worldsheets exponentiates these fields
into the \sigmod\ action \eq{sigmodaction}. Note however the effect of
the sum over spin structures--because of the insertions of arbitrary
numbers of $RR$ operators, it is no longer the case that the partition
function can (necessarily\footnote{There may exist special cases where
this indeed does occur.}) be represented in the form \eq{spinstruct}.
This fact is consistent with the absence of worldsheet supersymmetry in
the \sigmod. The superconformal symmetry of the flat spacetime \sigmod\
is broken to conformal symmetry by the $RR$ background. There are
several puzzles associated with this fact that we would like to consider
here. First, the flat spacetime theory has a superghost conformal field
theory, and one may ask the significance of this when worldsheet
supersymmetry is broken. In the present formulation, this ghost sector
becomes irrevocably tied up with the matter CFT. A related issue is the
appearance of picture. Recall first the source of picture in the flat
spacetime theory. The $\beta\gamma$ system is bosonized in terms of
fields $\eta,\xi,\phi$, and the field $\xi$ has a zero mode on the
sphere. One must then include a factor of $\xi$ in the string path
integral; BRST invariance is guaranteed if the operator
$P_{+1}=\{Q_{BRST},\xi\}$ is independent of $\xi$. We will argue in what
follows that these properties persist in Ramond-Ramond backgrounds:
although the detailed forms of $Q_{BRST}$ and $P_{+1}$ are modified,
there is still a $\xi$ zero mode, and hence the picture structure is
maintained.

\section{Contact terms}\label{sec:contact}

Contact terms in string theory are required for various reasons. As
Green and Seiberg \cite{GS} pointed out, contact terms arise in the OPE
of vertex operators. Contact terms are required in order to cancel
contributions to S-matrix elements due to unphysical states, and
therefore they serve to ensure the analyticity of the S-matrix and
worldsheet supersymmetry. In some other cases contact terms are required
to preserve spacetime supersymmetry. Alternatively,\cite{DD} contact
terms may be thought of in terms of our inability to enforce the
superconformal gauge globally when operators collide.

Contact terms also arise \cite{Kutasov} when one studies the moduli
space of a family of conformal field theories. In this case they
represent the connection coefficients of the metric in moduli space. As
the connection depends on a choice of coordinates, contact terms are not
uniquely determined at a given point in moduli space.

Regardless, it is clear that to properly define the theory, we have to
give, in addition to the field content and \sigmod\ action, a
prescription for contact terms. In the backgrounds considered in this
paper, this is non-trivial, since worldsheet supersymmetry is not
present in the sense discussed above. Instead, the appropriate
prescription is to enforce spacetime symmetries.

As an example, let us consider deforming a string theory compactified on
a torus by the $(1,1)$ operator $\varepsilon_{ij} \int d^2z\ \partial
X^i \bar\partial X^j$, which corresponds to a graviton at zero momentum.
The propagator $X^i(z) X^j(0)\sim \delta^{ij}\ln|z-z'|^2$ implies that
there results a quadratic divergence proportional to
$\varepsilon_{ij}\delta^{ij}\delta^{(2)}(0)$. This may be cancelled by
extending the vertex operator to an integrated superfield
\beq
V = \varepsilon_{ij} \int d^2 z \left( \pa X^i\bar\pa X^j
+\psi^i\bar\pa\psi^j+\bar\psi^i\pa\bar\psi^j+F^i F^j\right).
\eeq
The additional terms are contact terms and they vanish on-shell,
$F^i=0=\bar\pa\psi^i$. However, in a given correlation function, they
potentially contribute when the insertion collides with other operators.
It is at these points thet one cannot use the equations of motion, and
the terms which are zero on-shell produce a contact term contribution to
amplitudes.

For general deformations $\phi^i(x)$ produced by $(1,1)$ operators on
the worldsheet we expect the following structure for the OPE
\beqn
\phi^i(x,z) \phi^j(x,z') &=& \frac {\Delta^{ij}(x)}{|z-z'|^4}+
\frac1{|z-z'|^2}\ O^{ij}_k\phi^k(x,z)\\&&
+\frac{b^{ij}}{|z-z'|^2}\delta^{(2)}(z-z')+ c^{ij}\delta^{(2)}(z-z')\\
&&+\delta^{(2)}(z-z')\ \Gamma^{ij}_k\phi^k(x,z)+
\hbox{higher order}
\eeqn
The higher order terms correspond to $\alpha'$ corrections to the OPE.
The power singularities will contribute a quadratic divergence on the
worldsheet (proportional to $\Delta^{ij}$) and a one loop
$\beta$-function of $\phi^k$, whereas the $\delta$-function contact
terms serve to cancel the quadratic divergences  by making them into
total worldsheet derivatives. If one keeps the calculation to one loop
order the contact terms also serve to modify propagators to give the
correct background field renormalization. The contact term OPE 
corresponds to
\begin{equation}
\phi^i(x,z) \phi^j(x,z')|_{ct} =\frac{b^{ij}}{|z-z'|^2}
\delta(z-z')+ c^{ij}\delta^2(z-z')+ \delta^{(2)}(z-z')\ \Gamma^{ij}_k\phi^k(x,z)
\end{equation}
and $b^{ij}, c^{ij}$ are related to $\Delta^{ij}$, so that when they are 
combined they form a total derivative on the worldsheet. This is how 
the tachyon decouples from the CFT.

One has to keep in mind that the coefficient $\Gamma^{ij}_k$ depends on
the background, and gets corrected order by order; so when one does a
set of calculations to a given order, one has to give a prescription for
these corrections. This is the coefficient which is interpreted as the
connection on the moduli space of CFT \cite{Kutasov}.

In the superspace formulation, cancellation of quadratic divergences is
automatic, and thus it is easier to enforce conformal invariance on the
worldsheet. Indeed, these are the contact terms which are required to
cancel contributions to the S-matrix elements by unphysical particles,
and they always show up as total derivatives on the worldsheet. One can,
on the other hand, integrate out the auxiliary fields, and one gets
explicit contact terms in the Lagrangian. In the non-linear \sigmod,
these contact terms serve to cancel the quadratic divergences on the
worldsheet, and they also serve to restore the worldsheet supersymmetry,
which is non-linearly realized. In the absence of worldsheet
supersymmetry, this is the only way to proceed, so at each order in the
perturbation one has to calculate the appropriate contact terms.

In this paper we wish to understand $RR$ backgrounds, so one has two
types of problems to address. First, one still needs contact terms to
cancel the divergences on the worldsheet field theory, but these are
more complicated due to the fact that the $RR$ backgrounds break
worldsheet supersymmetry, and thus the superspace formulation is
ill-defined. One could hope that the contact terms introduced by the
$RR$ background are all zero, and that one could use the superspace
formulation for the $NS$ operators and thus obtain a result which is
free of quadratic divergences. Indeed, as we will see later, this is
precisely what happens in the $\alpha'$ expansion, but only to one loop
order.

Secondly, contact terms for theories with $RR$ backgrounds are subtle
because of picture. Given a prescription for contact terms in a fixed
picture, there could arise additional contact terms from
picture-changing operators. This would imply then that contact terms are
picture dependent. Fixing the $RR$ operators to be in picture zero (by
introducing the square root of the picture-changing operator) should
remove this ambiguity
% and therefore it can be used to 
%fix the contact terms for the $RR$ operators in a (picture independent) 
%fashion.

\section{One-Loop \sigmod\ Calculations}
\label{sec:oneloop}

Let us now demonstrate some features of \sigmod\ perturbation theory. We
do this for several reasons, mainly to demonstrate that the technique is
straightforward and systematic. We begin with one-loop computations, and
comment on higher order calculations later in the paper. As mentioned
previously, we assume for simplicity throughout that the dilaton is
constant, and the $NS$ $B$ field is unexcited. The $NSNS$ part of the action
is
\beq
I_{NSNS}={1\over2}\int d^2x\ \left\{ g_{ij}(\phi) \pa^\mu X^i \pa_\mu X^j
+ig_{ij}(\phi)\bar\psi^i (\Dsl\psi)^j+{1\over6}R_{ijk\ell} (\bar\psi^i\psi^k)
(\bar\psi^j\psi^\ell)\right\}
\eeq
A choice of contact terms must be made at lowest order. As we have
discussed, it is consistent (at least at leading orders) to define these
as if the $NSNS$ part of the theory possessed (off-shell) worldsheet
supersymmetry.
\beq
I_{NSNS}={1\over 4i}\int\ d^2x\ d^2\theta\ g_{ij}(\sX) \bar \calD\sX^i \calD\sX^j
\eeq
As is well-known, this choice cancels all quadratic divergences at one
loop. Alternate (on-shell) contact term prescriptions also exist, but we
find the present formulation most convenient computationally.

The $RR$ part of the \sigmod\ is written
\beq
I_{RR}=\int d^2z\ {\cal S}^\alpha \tilde {\cal S}^\beta h_{\alpha\beta}(X)
\eeq
where ${\cal S}^\alpha$ are spin fields
\beq
{\cal S}^\alpha=\phf S^\alpha e^{-\phi/2}
\eeq
To define $S^\alpha$, we introduce an auxiliary orthonormal frame
$e^\mu_a$, such that $\psi^\mu = e^\mu_a \chi^a$. The $\chi^a$ fermions
are conventionally normalized and may be bosonized in the standard
fashion. The spin fields $S^\alpha$ are then defined in the usual way,
mutual locality restricting them to one chirality. These are spinors of
the spacetime manifold, and the monodromies introduced by $S^\alpha$ on
the worldsheet fermions are also carried by the superghost
system. $\phf$ and it's right moving counterpart will be defined
implicitly by their square, which is also formally $P_{+1}= \{ Q_{BRST},
\xi\}$. The ghost system is the standard one,\cite{FMS} with $\gamma=
e^{\phi}\eta$, $\beta=e^{-\phi}\xi$, where $\eta,\xi$ are fermions of
conformal dimensions one and zero respectively. The point here is simply
that in the local frame, the ghost and picture-changing structure is
identical to the flat spacetime quantities. We will argue that in fact
the picture -changing operator receives no further corrections in
\sigmod\ perturbation theory, at least to low orders.

Relevant OPE's may be written
\beq
{\cal S}^\alpha(z)\ {\cal S}^\beta(0)\sim 
{(\Gamma C^{-1}\cdot\pa X)^{\alpha\beta}\over z}
\eeq
\beq
\chi^a(z)\ S^\alpha(z') \sim {(\Gamma^a S)^\alpha\over (z-z')^{1/2}}
\eeq
\beq
:\chi^a\chi^b:(z)\ S^\alpha(z') \sim {([\Gamma^b,\Gamma^a]S)^\alpha
\over z-z'}
\eeq
For the calculations which follow, we will employ a normal coordinate
expansion (see Ref. \cite{GFM} for details).

\FIGURE{\littlefig{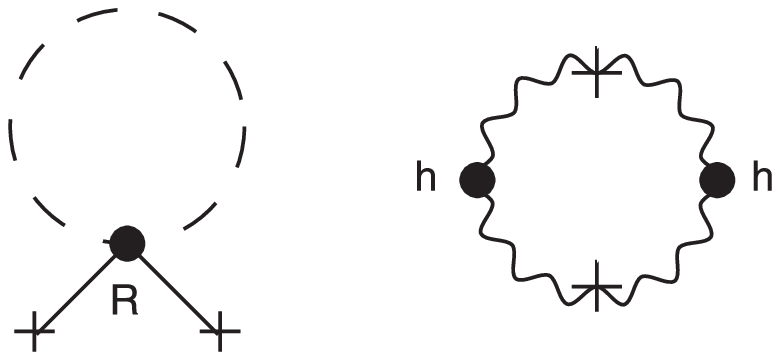}{2}
\caption{The one-loop graph from $NSNS$ 
and $RR$ sectors, respectively. Wavy lines
denote ${\cal S}$ propagators.}}

Let us consider the graviton $\beta$-function. At one loop, there are
two contributions linear in curvature.\footnote{Two powers of the field
strength $h$ count consistently as a single power of curvature.}
From the $NSNS$ graph, we obtain
\beqn
{1\over2}\int {d^2z\over 2i}\ R_{ik_1k_2j}\pa X^i\bar\pa X^j e^{k_1}_a(X)
e^{k_2}_b(X)\langle \xi^a(z)\xi^b(z)\rangle\nonumber\\
={1\over 4}\Delta(\epsilon)\int d^2z\ R_{ij} \pa X^i\bar\pa X^j\nonumber
\eeqn
where $\xi^a$ are canonically normalized bosonic fluctuations, with
$\langle \xi^a(z)\xi^b(z')\rangle=\eta^{ab}\Delta(z-z')$.

From the $RR$ graph we find
\beqn
{1\over 2!}\int d^2zd^2z'\ {\cal S}^\alpha \tilde {\cal S}^\beta
h_{\alpha\beta}(z)\cdot {\cal S}^\gamma 
\tilde {\cal S}^\delta h_{\gamma\delta}(z')\nonumber\\
={1\over 2!}\int d^2z\int {d^2z'\over |z-z'|^2}\ \pa X^i\bar\pa X^j 
e^{a}_i(X)e^{b}_j(X)
\tr \left( h\Gamma_a C^{-1} (\Gamma_bC^{-1} h)^T\right)
\eeqn
To evaluate this further, let us specialize to the case of a single
$p$-form field strength
\beq
h_{\alpha\beta}={1\over p!} H_{\mu_1\ldots \mu_p} 
\left( C\Gamma^{\mu_1\ldots \mu_p}\right)_{\alpha\beta}
\eeq
We need then evaluate
\beq
\tr \left( \Gamma^{\mu_1\ldots \mu_p}\Gamma_a\Gamma^{\nu_p\ldots \nu_1}
\Gamma_b\right)H_{\mu_1\ldots \mu_p}H_{\nu_1\ldots \nu_p}
=32\ p! \left( H^2_{ab}-{1\over 2p} \eta_{ab} H^2\right)
\eeq
Thus, we may write the $\beta$-function at this order as
\beql{fullbetafunction}
\beta_{ij}=R_{ij}-{1\over 2 (p-1)!}\left( H_{ij}^2-{1\over 2p}g_{ij}
H^2\right)
\eeq
Note that this is the correct equation of motion for constant dilaton
(for which $R=0$; in the case of $AdS_p\times S_p$, this is achieved by
cancellation between the two factors.)

We have computed the $\beta$-function to the operator of order $\xi^2$; 
there are of course also $\beta$-functions for terms of higher order. 
These however may be interpreted geometrically as the normal coordinate
expansion of the curvature tensors appearing in the above $\beta$-function.
There are tree-level corrections to the
picture-changing operator itself, which come only from the $NSNS$ background (we
discuss this fact more fully later in the paper). The modification simply
involves a covariantization $\eta_{\mu\nu}\psi^\mu\pa X^\nu\to 
g_{\mu\nu}(X)\psi^\mu\pa X^\nu$. This is simply related to the normalization
of the field propagators; the spin field itself however is given directly
in an orthonormal frame (see Appendix), and so does not undergo such renormalizations
in the $NSNS$ background.

Let us make a few comments concerning the contraction of the two $RR$
operators. At lowest order, there are no contact terms, and the
divergence is as given. In particular, the operators $\phf$ and $\bphf$
are harmless; there are two of each, and at this order, they simply
combine to $P_{+1}\bar P_{+1}$. At higher orders, there is a possible
contact term between $\phf$ and $\bphf$ (as there would be between
$P_{+1}$ and $\bar P_{+1}$), and this needs to be taken into account
consistently. We will return to this discussion in the next section.

\section{BRST}\label{sec:brst}

Let us turn our attention now to the structure of BRST in theories with
RR backgrounds. In order to emphasize the various features, we begin
with a short review of standard material for the ordinary superstring.
This discussion will also serve to set notation. For a more complete
discussion we refer the reader to \cite{GSW,Joe}.

Traditional string compactifications are described in terms of a
worldsheet supersymmetric \sigmod\ coupled to worldsheet supergravity.
The full theory is superconformally invariant. For simplicity of the
discussion we will assume that the dilaton is constant (in this case the
theory is also classically superconformally invariant).

Locally on the worldsheet one can  choose the superconformal gauge and
in this gauge one has eliminated the local worldsheet gravity degrees of
freedom. The supergravity ghost system in this gauge is conformally
invariant,  and is described by a free field theory which is decoupled
from the matter superconformal field theory.

The metric and gravitino enter in the action as Lagrange multipliers,
and their equations of motion imply that the (super) stress tensor of
the system vanishes. Classically one has $T_{z\bar z}=0$ (because the
theory is conformally invariant). The holomorphic and anti-holomorphic
pieces of the matter stress tensor do not vanish as operators, and 
they can only be made to vanish 
%as operator equations. 
on states. Because there is a conformal anomaly for the matter system,
the stress tensor can not be made to vanish identically, independent of
the ghost system. Solving the constraints is done via the BRST
quantization of the string theory.

The string theory is well posed (the target space \sigmod\ is a solution
of string theory) if the total central charge of the conformal field
theory of the combined matter and superghost system vanishes. The ghost
system is described by two conjugate superfields $B=\beta+\theta b$ and
$C=c+\theta \gamma$, of dimensions $3/2$ and $-1$. In this case,  one
has an (anti)holomorphic BRST operator $Q$ which squares to zero, and is
described by the superfield
\begin{equation}
Q = \oint dz d\theta (C T_{Gm} +\frac 12 C T_{Gg})\label{eq:BRST1}
\end{equation}
where $T_{Gm}=G_m+\theta T_m$ is the super-stress tensor of the matter
system and $T_g= G_g+\theta T_g$ is the stress tensor of the decoupled
ghost system \footnote{Notice that for conformal field theories with
more than one worldsheet supersymmetry we actually have a choice as to
which supersymmetry we use to define $G_m$. All of these choices should
be unitarily equivalent.}.

Because the ghost system is free, one has two separately conserved ghost
numbers (one for the $b,c$ system, and another for the $\beta,\gamma$
system), and the BRST operator splits into pieces of different fermionic
ghost charge, namely
\begin{equation}
Q = Q_0 +Q_1+Q_2
\end{equation}
where
\begin{eqnarray}
Q_0&=&\oint c (T_m+T_{\beta\gamma} +\frac 12 T_{bc})\\
Q_1&=&\oint \gamma(G_m)\\
Q_2&=&\oint \frac 12 b\gamma^2 \label{eq:BRST2}
\end{eqnarray}
$Q^2=0$ gives independent equations for each ghost number, so we get
$Q_0^2=0$, $Q_1^2= -\{Q_0,Q_2\}$ and $Q_2^2=0$. The first of these
equations is the nilpotency of the standard conformal BRST charge, which
says that the superconformal field theory is conformally invariant.

Physical states are associated with the cohomology of $Q$. To each state
one associates a vertex operator $\CO_i$ which is superconformally
invariant, $Q \CO_i =0$. This solves the constraints produced by the
equations of motion of the worldsheet supergraviton. The constraints put
the physical state on-shell in the spacetime theory.

The operators are given with fixed ghost charges. For the
$b,c$ system, unintegrated vertex operators are built as
$\CO_i = c \bar c V_{1,1}$, where the $V^{(i)}_{1,1}$ is a vertex
operator of conformal dimension $(1,1)$ that is independent of 
$b,c$ (but which might depend on $\beta,\gamma$). 
The integrated operators are obtained by removing the 
$c\bar c$ and replacing it by $\int d^2 z$.
$Q \CO_i=0$ implies that $Q_0 \CO_i=0$, so the operator is 
also conformally invariant. Correlation functions are
constructed by integrating the vertex operators $\CO_i$ over a
Riemann surface with fixed complex structure; and then integrating
over the moduli space of complex structures (these are the 
non-local degrees of freedom that the superconformal 
gauge does not remove globally). Because of the worldsheet fermions
one also needs to pick a spin structure on the Riemann surface.

The different parts of the BRST charge play different roles. Indeed,
vertex operators are conformally invariant if $Q_0 V=0$. The extra
pieces of the BRST operator give additional constraints on the system.
It is particularly important to notice that $Q_0$ acts as a second order
differential operator in spacetime, so it is an appropriate tool to give
equations of motion to spacetime bosons.

For spacetime fermions, $Q_0$ gives a Laplace equation, with double the
appropriate number of degrees of freedom. The extra pieces of the BRST
operator give an additional constraint, namely the Dirac equation.

Notice that one can interpret this fact in a slightly different way.
First, ignoring the superconformal invariance, one could just have
required conformal invariance and say that supersymmetry was accidental.
This is how one builds the bosonic string in the absence of
supersymmetry after all. In the special case where the theory is
supersymmetric, there is an extra operator $Q_1$ which squares to zero
in the $Q_0$ BRST cohomology classes, $(Q_1)^2 = - Q_{0} Q_2
=0$. Thus, in the cohomology of the $Q_0$ BRST operator, $Q_1$ acts as
the $d$ operator of an exact sequence (graded by the ghost charge), so
one can use it to give further restrictions on cohomology (the
idea behind spectral sequences). From a spacetime point of view, we are
getting a first order differential operator acting on states.

\subsection{BRST for $RR$ backgrounds}

In the calculations presented so far,  the picture-changing operator for
RR backgrounds was defined without any reference to a BRST structure on
the worldsheet, indeed we used the standard picture-changing operator.
The purpose of this section is to give a (somewhat heuristic)
description of BRST that justifies our comments; this should serve
as a basis for a full formal perturbation expansion to all orders.

It is instructive to analyze how the BRST charge gets modified as we
change the background in which the string propagates. We have a family
of conformal field theories (parametrized by the perturbation
parameters), and the BRST operator is a section of an operator bundle
over the moduli space of conformal field theories.

Let us begin with a supersymmetric \sigmod\ parametrized by
a metric on target space $g^{(\alpha)}_{\mu\nu}(x)$; 
the stress tensor is given by
\begin{equation}
T= \frac 12 g^{(\alpha)}_{\mu\nu}(x)\partial X^\mu\partial X^\nu +
\hbox{fermions}
\end{equation}
with a similar expression for the supersymmetry generator. As we change
$g$, the stress tensor changes appropriately, at tree level.

As the $BRST$ charge is intimately tied to the stress tensor,
we expect that it is modified just as the stress tensor is.
Indeed
\begin{equation}
\delta Q_{BRST} = \delta \ C T_{Gm}
\end{equation}

We want to justify this answer directly from the perturbative approach
to the family of conformal field theories. Indeed, the tangent space in
the moduli space is generated locally by the massless $(1,1)$ vertex
operators, which are BRST invariant, so it does not seem necessary to
modify the BRST operator locally on the moduli space.

However, there is a caveat in the argument, which stems from the fact
that the perturbations are done by integrated vertex operators. When one
is checking the invariance of the background one needs to make sure to
take into account the contact terms between the integrated vertex
operator and the BRST contour. These contact terms need to be cancelled,
as they would spoil the BRST invariance of the background, and they
provide the necessary corrections to the BRST charge.

The fact that to first order one can ignore contact terms is
justified by how one parametrizes moduli space. Indeed, the contact
terms act as the connection on the operator bundle on the moduli space
of conformal field theories \cite{Kutasov}. One can choose a local
parameterization of the moduli space in which the connection vanishes,
which is also the reason why contact terms are not fully determined when
one considers string amplitudes. In spacetime this choice is
characterized by the different gauge choices in the supergravity low
energy effective field theory.

The BRST operator, being essential to the quantization of the string,
should be a canonical object in this moduli space of backgrounds. One
can construct the BRST operator by following a path from our starting
point (flat space in our case) to another SCFT in the same moduli space.
The BRST operator of the perturbed theory is constructed by parallel
transport along the path. As there might be monodromies in the bundle
over moduli space, the best one can hope for is that given two different
paths between the same points in moduli space, the two BRST operators
give string theories which are unitarily equivalent. It is in this sense
that the BRST operator associated to a point in moduli space is
canonical.

For $RR$ backgrounds, we will take the approach that the BRST operator is
defined by  parallel transport of the BRST operator in the $NS$ conformal
field theories on the path that connects it to the point we are
studying.

From this construction, and the fact that the physical states can always
be taken to be independent of the $\xi$ zero modes coming from the
$\beta\gamma$ ghost system, we learn that these zero modes survive even
in the presence of the $RR$ perturbation of the CFT. The BRST operator
thus obtained will give us a picture-changing operator $P_{+1}= \{
Q_{BRST},\xi\}$.

Now, we will  argue that there is no contact term that makes the
picture-changing operator depend on the $RR$ field strength at tree level
on the worldsheet, as we have assumed in Section \ref{sec:oneloop}. To
analyze this question we will concentrate on the $\eta$-dependent part
of the BRST operator, $Q_1$. Since we are considering a perturbation
expansion, we indicate the expansion order in the BRST current by an
upper index; the lower index continues to refer to the ghost charge.
Thus, if we write the background vertex corresponding to the full
deformation as
\begin{equation}
V_B = \sum_{i=1}^{\infty} \epsilon^i V_{(i)}
\end{equation}
where $\epsilon$ is the formal perturbation expansion parameter (in
normal coordinates the expansion parameter is the  coordinate itself),
we also write
\begin{equation}
Q_j = \sum_i \epsilon^i Q_j^{(i)}
\end{equation}
for the BRST charge. $Q_i^{(0)}$ are the pieces of the unperturbed BRST
charge. For our purposes, we will only analyze the first order
corrections to $Q$, namely $Q_0^{(1)}$ and $Q_1^{(1)}$ which are the corrections
coming from the $RR$ vertex operator. $Q_0^{(2)}$ is determined by the contact
term between a $NS$ operator and the $Q_0$ BRST current (a standard
computation).

Contact terms usually arise because a left moving field and a right
moving field are located at the same point, e.g.
\begin{equation}
\partial X^\mu(z) \bar\partial X^\nu(z') \sim \delta^{(2)}(z-z') \eta^{\mu\nu}
\end{equation}
in a free field theory.
It is easy to see that these type of terms are precisely the ones that
make modifications to the BRST operator in the $NS$ backgrounds. So let us
consider the left moving BRST operator (just the piece with ghost charge
1), and the insertion of a $RR$ background vertex. We want to analyze a
possible contact term between $Q_1^{(0)}\sim\gamma G_m$ and the vertex.

Now, we would like to be able to directly calculate this effect with $V_{RR}$
in the $(0,0)$-picture ({\it i.e.,} with the factors of $\phf$ present).
However, one must be careful, because there could be contact terms with
the $\phf$'s. Indeed, we would like to motivate that there is definitely
such a contact term present. To do so, we will consider the $RR$ background
insertion in various pictures, and show that the resulting contact terms
are picture-dependent.\footnote{Here we specifically mean that results
are different if we move picture-changing operators to be coincident
with the background insertion.}  This ambiguity is removed if we agree to take
the background insertions always in picture $(0,0)$. In a later section,
we will work out the contact terms in that picture, using as a guiding
principle, worldsheet conformal invariance.

As a first step, which justifies the treatment at one-loop that we gave
in Section \ref{sec:oneloop}, we argue that this ambiguity is not
present at tree-level -- it is a one-loop effect. 

If one uses a $(-1/2,-1/2)$-picture operator, there seems to be no
apparent contact term, because the left and right moving fermion field
theories are decoupled. On the other hand, suppose we consider a picture
$(-1/2,1/2)$ vertex operator, $h_{\alpha\beta} (\Gamma_\mu)^\alpha_{\
\dot\beta}\ \tiden\beta\bar\pa X^\mu\ S^\beta e^{-\phi/2}$. Here, we have
defined
\beq
\iden\alpha\equiv S^{\dot\alpha}e^{+\phi/2},\ \ \ \ 
\tiden\alpha\equiv \tilde S^{\dot\alpha}e^{+\bar\phi/2}
\eeq 
There is  a contact
term coming from the $\partial X^\mu$ in $G_m$ and the $\bar\partial
X^\mu$ in the vertex. Also, the $\gamma\psi^\mu$ of $Q_1^{(0)}$
has an OPE with the spin fields. The resulting contact term counts as a
1-loop effect, so we learn that the tree level correction to the
picture-changing operator vanishes. Thus, we have justified our claims
regarding the effects of picture-changing in the one-loop computations
of Section \ref{sec:oneloop}. In particular, since the $RR$ correction to
picture changing vanishes at tree level, the picture-changing operator
acts, in one-loop computations, as if there were a supersymmetric
\sigmod. Indeed, the one loop result is globally well-defined.

If one works out the details in picture $(-1/2,1/2)$, one finds that the
the contact term is equivalent to a one-loop correction to the BRST operator takes the form\footnote{Here,
$h_{\dot\alpha\dot\beta}\sim h_{\alpha\beta}(\Gamma_\mu)^\alpha_{\ \dot\alpha}
(\Gamma^\mu)^\beta_{\ \dot\beta}$.}
\beq
Q_1^{(1)}\sim
\eta\ \iden\alpha\tiden\beta
h_{\dot\alpha\dot\beta}
\eeq
which is seen to have ghost charge $(1/2,1/2)$. The fact that the
contact terms are picture dependent, shows that there are ambiguities in
the definition of the contact terms. Note that the operator
$\iden\alpha\equiv S^{\dot\alpha}e^{+\phi/2}$
is of conformal dimension $(0,0)$.  However, since the CFT is not
unitary, it is not the case that $\iden\alpha$ is trivial. Indeed it has
non-trivial OPE's with itself and other operators.

Note that $Q_1^{(1)}\sim \eta\ h_{\dot\alpha\dot\beta}\iden\alpha\tiden\beta$
is not holomorphic. This is certainly a problematic feature. However, we
remind the reader that this explicit form was derived starting with a
picture $(-1/2,1/2)$ background insertion. It's presence is meant to
merely indicate the necessity of having a contact term. The fact that it
is non-holomorphic, we believe, may be traced to the fact that the
superstring is not really a free field theory (rather, it only appears
so if one ignores picture). In the next section, we will give a
holomorphic prescription for the contact term (for picture $(0,0)$
backgrounds) which has the right structure.

That this switch to a holomorphic form is necessary is clearly indicated
by looking at the contact term related to $Q_0^{(1)}$, {\it i.e.}, the
term proportional to $c$. Again, it is picture-dependent and if we
choose the $RR$ background in $(-1/2,1/2)$ picture, we find
\beql{RRcontact}
\sim\oint dz\ c\ h_{\alpha\beta} (\Gamma_\mu)^\beta_{\ \dot\beta} \
\CS^\alpha \pa X^\mu \ \tiden\beta
\end{equation}
This correction may be interpreted as a correction to $T_{zz}$, and so
must be holomorphic by conformal invariance. Again, we will find the
proper resolution in the next section.

\subsection{Background BRST invariance}

When the background is taken in $(0,0)$ picture, the correction to the
BRST operator may be determined unambiguously. We can determine this by
requiring that the one-loop $\beta$-functions are recovered explicitly
through BRST invariance of the background. In particular, a tree-level
correction to the BRST operator, proportional to the $RR$ fieldstrength,
is required.

As a warmup, let us recall how this works in a purely gravitational $NSNS$
background. The bosonic term in the \sigmod\ is at second order
proportional to
\beql{secondorderNS}
R_{\mu\nu\rho\sigma} X^\mu X^\rho\pa X^\nu \bar\pa X^\sigma
\eeq
and the correction to the worldsheet stress tensor results from
exchanging $\bar\partial X^{\sigma}\to \partial X^{\sigma}$. Note that
the resulting operator is holomorphic on-shell at this order in $R$.

The BRST current applied to the correction to the background gives
\beql{secondorderfull}
Q^{(0)} \cdot\int d^2 z\ V_{(1)} \sim \int d^2 z\ \pa c \cdot
R_{\mu\nu}\partial X^\mu\bar\partial X^\nu
+\hbox{total derivative}
\eeq
Requiring that this vanishes is completely equivalent to the vanishing
of the graviton $\beta$-function.

 \bigskip
 
Now, we would like to analyze this with the $RR$ background. In order to
see an effect, we need to expand to second order in the background.
To this order then, we want to test
\beql{confinv}
Q_0^{(0)}\cdot\left(-\int d^2z\ V_{(2)} +\frac 12(\int d^2z\ V_{(1)})^2 
\right) -Q_0^{(1)}\cdot V_{(1)} =0 
\eeq
Each vertex is meant to be inside the contour of $Q$. The factors of
$V_{(1)}$ are $RR$ background insertions, while $V_{(2)}$ is the $NSNS$
quantity given in \eq{secondorderNS}. $Q_0^{(1)} = Q_0^{(0)}
V_{(1)}|_{c.t}$ is the contact term between the original BRST operator
and the $RR$ background.

We work to one-loop. The term involving $(\int V_{(1)})^2$ may be
dropped because its contribution either vanishes on-shell (from
$Q_0^{(0)}\cdot V_{(1)}=0$) or is of higher loop order (a contact
contribution from $V_{(1)}V_{(1)}\sim \delta$ is already of one-loop
order).

The result of $Q_0^{(0)}\cdot V_{(2)}$ is as calculated above for the
NSNS background, and results in a contribution to the graviton
$\beta$-function $\sim R_{\mu\nu}\partial X^{\mu}\bar\partial X^{\nu}$.

Next, we need to calculate $Q_0^{(1)}\cdot V_{(1)}$. Previously, we 
considered $V_{(1)}$ in picture $(-1/2,1/2)$ and found the result
\beql{nonholoQ}
Q_0^{(1)}\sim\oint dz\ c\ h_{\alpha\beta} 
(\Gamma_\mu)^\beta_{\ \dot\beta}\
e^{-\phi/2}S^\alpha \partial X^\mu \ \tiden\beta 
\eeq
We would now like to find a suitable $Q_0^{(1)}$ when $V_{(1)}$ is
in picture $(0,0)$, such that eq. \eq{confinv} is satisfied. This will
uniquely fix $Q_0^{(1)}$, including its coefficient.

Let us demonstrate that eq. \eq{nonholoQ} is problematic if we use it literally,
including appropriately chosen picture-changing operators to bring the
full result to $(0,0)$ picture. Indeed there is a non-holomorphic
singularity present
\beq
\oint dz\ \frac 1{|z-z'|^2} c\ h_{\alpha\beta}h_{\gamma\delta} 
\partial X^\mu \partial X^\nu 
(\Gamma_\mu)^{\beta\delta}(\Gamma_\nu)^{\alpha\gamma}
\eeq
To avoid this, we choose a contact term, i.e., a new version of $Q_0^{(1)}$
\beq
Q^{(1)}_0 \sim \oint dz\ c\ h_{\alpha\beta} \CS^\alpha \Sigma^\beta
\eeq
If we require the OPE
\beq
\Sigma^\alpha(z) \tilde\CS^\beta(z') \sim \frac{1}{z-z'}\bar\pa X^\mu 
(\Gamma_\mu)^{\alpha\beta}
\eeq
then we obtain
\beq
Q^{(1)}_0\cdot V_{(1)}\sim (hh)_{\mu\nu}\pa X^\mu\bar\pa X^\nu
\eeq
The tensor form is precisely right such that 
 eq. \eq{secondorderfull} reduces to the full graviton $\beta$-function,
\eq{fullbetafunction}.

\bigskip

Next, note that the above form for $Q_0^{(1)}$ could be deduced via
a contact term calculation if we have also have 
\beqn
\left.T(z)\cdot \tilde\CS^\beta(\bar z')\right|_{c.t.}
&\sim& \delta^{(2)}(z-z')\Sigma^\beta(z')\\
\left.T(z)\cdot \CS^\beta(z')\right|_{c.t.}&\sim& 0
\eeqn
Furthermore, note that we may equivalently write a correction to the
stress tensor from the $RR$ background of the form
\beq
T^{(1)}\sim h_{\alpha\beta}\CS^\alpha\Sigma^\beta
\eeq
The stress tensor is thus properly holomorphic, and has holomorphic
OPE's with other operators.

\subsection{BRST and Physical States}

In our previous paper, we demonstrated mixing between $NS$ and $R$ operators
in the presence of $RR$ backgrounds. We can no show, with our prescription
for $Q_0^{(1)}$, that those results can be consistently reproduced via
BRST methods. Indeed, this is an important check of our methods.

As an example, consider the massless case. Physical vertex operators
in unintegrated form are
\beq
\CO_{NS} = P_{+1}\bar P_{+1}c\bar c\ f_{\mu\nu}e^{ikx}\
\psi^\mu e^{-\phi}\ \bar\psi^\nu e^{-\bar\phi} 
\eeq
and 
\beq
\CO_{RR}= \phf\bphf c\bar c\ g_{\gamma\delta}e^{ikx}\ 
e^{-\phi/2} S^\gamma\   
e^{-\bar\phi/2}\tilde S^\delta 
\eeq
Consider a state which, at zeroth order, is a $RR$ state. We know that this
mixes with an $NSNS$ state, and so the condition, at first order, for the
full state ${\cal O}_{RR}+\epsilon {\cal O}_{NSNS}$ to be physical is
\beq
Q_0^{(0)}{\cal O}_{NSNS}+Q_0^{(1)}{\cal O}_{RR}=0
\eeq
To check this, we need only $Q_0^{(1)}$, which we have given above. 
Performing the OPE's, we find
\beq
c\pa c\bar c\psi^\mu e^{-\phi}\tilde\psi^\nu e^{-\bar\phi}
\left( k^2f_{\mu\nu}+h_{\alpha\beta}g_{\gamma\delta}
(\Gamma_{\mu})^{\alpha\gamma}(\Gamma_{\nu})^{\beta\delta}\right)
\eeq
This is precisely of the correct form; its vanishing determines the
appropriate mixing (see Ref. \cite{first} or \cite{VanN}).

Next, if we want to demonstrate the mixing for a physical state of
the form ${\cal O}_{NSNS}+\epsilon {\cal O}_{RR}$, we need a 
prescription for $Q_1^{(1)}$:
\beql{eqom}
Q_1^{(1)}{\cal O}_{NSNS}+Q_1^{(0)}{\cal O}_{RR}=0
\eeq
Recall that previously we had $Q_1^{(1)}\sim\eta h_{\dot\alpha\dot\beta}
\iden\alpha\tiden\beta$. We wish to replace this by a holomorphic
expression. To do so, introduce a new field $\Sigma^{\dot\alpha}$ with OPE
\beq
\Sigma^{\dot\alpha} \bar\pa X^\mu \sim {1\over z-z'} 
(\Gamma^\mu)_\alpha^{\ \dot\alpha}
\tilde\CS^\alpha
\eeq
Then, the choice\comment{Added oint}
\beq
Q_1^{(1)} \sim a\oint{\rm P}_{-1/2} \eta e^{3\phi/2} S^\alpha\pa X^\mu
(\Gamma_\mu)_\alpha^{\ \dot\alpha} h_{\dot\alpha\dot\beta}
\Sigma^{\dot\beta}
\eeq
is holomorphic and of the appropriate dimension and picture. $a$ is
a normalization constant.
Eq. \eq{eqom} gives
\beq
\phf\bphf\eta c\bar c \iden{\alpha}\left(\ksl g+a\Gamma^\lambda\ksl h \Gamma^\nu
f_{\nu\lambda}\right)_{\dot\alpha\alpha}e^{-\bar\phi/2}\tilde S^\alpha e^{ikx}
\eeq
One can think of the extra term as arising from the variation of the
covariant derivative of the background; there is a choice of the normalization
constant $a$ which makes this consistent with spacetime supergravity.
This equation can be solved to find the required mixing for the physical
state.

\subsection {Comments}

In the previous sections on BRST, we have noted that in order to
obtain consistent results, we must modify the BRST charge. In order
for this to retain a holomorphic structure, we are forced to introduce
new fields, with somewhat `exotic' OPE's. Let us review the structure
of these fields. We have fields $\Sigma^\alpha$ and $\Sigma^{\dot\alpha}$
of dimension $(1,0)$. (Fields of
dimension $(0,1)$ would also be required for the antiholomorphic
part of the BRST charge.) These fields intertwine the matter CFT's via
\beqn
\Sigma^\alpha(z) \tilde\CS^\beta(z') &\sim& \frac{1}{z-z'}\bar\pa X^\mu(z') 
(\Gamma_\mu)^{\alpha\beta}\\
\Sigma^{\dot\alpha}(z) \bar\pa X^\mu(z') &\sim& {1\over z-z'} 
\tilde\CS^\alpha(z')(\Gamma^\mu)_\alpha^{\ \dot\alpha}
\eeqn
Note that such a structure,  mixing holomorphic and anti-holomorphic
quantities, is allowed here because the CFT is not
unitary. The alternative would be to discard BRST entirely.
Although this system has a simple structure, we caution that it is
by no means clear that it is complete. We have not for example checked
that the operator algebra closes.

Let us also comment further on the structure of the ghost CFT's. Since 
we have enforced conformal invariance ($T_{z\bar z}=0$), a conformal
gauge choice is possible, at least locally. Thus the conformal ghost
CFT is unmodified. The superconformal ghosts are another matter. We
have seen that although some of the structure, such as picture, is
retained in the full theory, although the ghost and matter CFT's are
mixed. Since worldsheet supersymmetry is lost in the $RR$ background, one
should simply think of the presence of $\eta,\xi,\phi$ as a parameterization
of the CFT which is convenient close to a trivial background.

Perhaps one puzzling aspect of the BRST analysis is the lack of worldsheet
supersymmetry. When supersymmetry is present, it provides a canonical 
square root for the spacetime Laplacian, related to $Q_1$, which squares to
zero in $Q_0$ cohomology. The existence of such an operator allows
for the appropriate constraints on spacetime fields.
Worldsheet supersymmetry is not a necessary
condition however, and it is possible to find an appropriate nilpotent
operator which leads to constraints consistent with spacetime supersymmetry.
Note that we have a theory with constraints, which 
are not derived from any apparent gauge principle. It is possible
of course that such a gauge principle exists.

\section{Two loops and higher order}\label{sec:twoloop}

In Section \ref{sec:oneloop}, we have obtained one-loop
$\beta$-functions in a normal coordinate expansion. In the last section,
we have seen that the picture-changing operator receives corrections
from the $NSNS$ background only at tree level-- it is covariantized with
respect to the full background metric. This means that there are no
subtleties for one-loop $\beta$-functions, and they are defined globally
as a power series in the normal coordinate expansion. The inclusion of
the $RR$ background led to no quadratic divergences, and there appeared to
be no need for contact terms of the $RR$ insertions. The BRST analysis
however led to the conclusion that contact terms may appear at higher
orders, and we were able to accommodate such terms effectively.

In fact, contact terms for $RR$ insertions are required by spacetime
supersymmetry at higher loop order. If there were no such contact terms,
then each pair of $RR$ insertions corresponds to an extra loop in the
$\alpha'$ expansion, and one obtains a logarithmic divergent factor for
each pair--as a result they do not contribute to the graviton
$\beta$-function. Thus, in the absence of $RR$ contact terms, the only
contribution to the graviton $\beta$-function would be of order $h^2$.
On the other hand, it is expected that at four loops in the $\alpha'$
expansion, the graviton $\beta$-function receives a correction of order
$R^4$. A spacetime supersymmetric completion would include terms with at
least four factors of the $RR$ fieldstrength $h$.

Contact terms for the $RR$ insertions can reduce the divergence of terms
of higher order in $h$ to a single logarithm, thus giving a contribution
to the $\beta$-function. Thus, such contact terms are required by
spacetime supersymmetry.

However, we do need to show that at low loop orders, there are
nevertheless no contributions to the $\beta$-functions, which is a 
generalization of the results of \cite{GFM}. Let us now
discuss the case of two loops. Consider the presence of three $RR$
background insertions--this would potentially give a contribution to the
$RR$ $\beta$-function. If we bring two of the insertions together, we
obtain a logarithmic divergence proportional to $R_{\mu\nu}\pa X^\mu
\bar\pa X^\nu$ (here we have, consistently, used the one-loop equations
of motion). To this, we should add
$R_{\mu\nu}(\psi^\mu\bar\pa\psi^\nu+\bar\psi^\nu \pa\bar\psi^\nu)$ The
resulting operator could have a contact term with the third $RR$
insertion. A short calculation reveals that this is proportional to
\beq
\delta^{(2)}(z-z') R_{\mu\nu} h_{\alpha\beta} 
(\Gamma^\mu\Gamma^\nu)^\alpha_\gamma
\CS^\gamma \tilde\CS^\beta
\eeq
Since $R_{\mu\nu}$ is symmetric, this expression simplifies, and the
$\beta$-function is proportional to $Rh_{\alpha\beta}$. However, since
we have restricted the backgrounds to constant dilaton, $R=0$. Thus,
there is no two-loop contribution to the $RR$ $\beta$-function from 
three $RR$ insertions. One should also take into account  a possible
contribution from loop corrections to the picture-changing operator.
However, one can again argue that this is proportional to $R$ and
thus vanishes for this special class of backgrounds,

Another possible source for the $RR$ $\beta$-function would be a single
$RR$ insertion plus terms of up to order $X^4$ in $NSNS$ fields. However,
again we find that all such contributions are proportional to the Ricci 
scalar, and so vanish.

We can also discuss the two-loop graviton $\beta$-function. From our
previous discussion, we know that the contribution of four $RR$
insertions would lead to a $(\log\epsilon)^2$ singularity in the 
absence of $RR$ contact terms. However, such contact terms are of one-loop
origin; a pair of such contact terms then counts as two-loops, and there
is no room left, at two loop order, for a further logarithmic divergence.
We conclude that a two-loop contribution of order $h^4$ to the graviton
$\beta$-function vanishes. 

Thus, all two-loop contributions to the $\beta$-functions vanish. These
results imply that there is a choice of contact terms which is compatible
with spacetime supersymmetry to this order. 
Indeed
for products of homogeneous spaces (like $AdS_5\times S^5$) it is easy
to see that the possible contact terms between the $NS$ and $RR$ field
vanish at one loop, again because they are proportional to
$R_{\mu\nu}g^{\mu\nu}$, and therefore there is no $\beta$-function to two
loops at all for the $RR$ fields.

Notice that in principle one can set the contact terms to zero up to two
loops for the $RR$ fields, and then there is no contribution from the $RR$
field to the three loop $\beta$ function, which is then consistent with
supergravity. If one introduces contact terms at three loops, they would
contribute to a four loop $\beta$ function for the graviton and the $RR$
fields. In spacetimes where one is supposed to have spacetime
supersymmetry, we have to worry that we can define the spacetime
supersymmetry current on the worldsheet consistently order by order in
the perturbation expansion. Indeed, contact terms should contribute to
show that the worldsheet symmetry current associated to supersymmetry
is conserved. This makes it hard to believe that the procedure of
setting all these contact terms to zero will work up to that high an
order. We do not wish however, to explicitly carry out the four loop 
computations.

\section{Conclusions and Comments}

In this paper we have further explored \sigmod\ perturbation theory
including $RR$ backgrounds. The expansion appears to be
consistent to all orders in $\alpha'$, with contact terms being
determined by {\it spacetime} supersymmetry. Worldsheet supersymmetry is
broken by these backgrounds; however, at least at low orders in
perturbation theory, it is possible to mimic the appropriate contact
terms by taking the $NSNS$ part of the \sigmod\ off-shell in a
supersymmetric form. This stems from the fact that the contact terms
between the $RR$ operators don't contribute at lowest orders for the
special class of backgrounds where the dilaton is constant.

Certain aspects of the theory remain quite simple. One has a notion of
BRST symmetry and the   BRST current is of  the form
\beq
J_{BRST}=J_{{\rm conf}}+J_1+J_2
\eeq
where $J_{{\rm conf}}$ is the contribution from the conformal field
theory stress tensor, and it annihilates physical states. The stress
tensor includes contributions from the $RR$ background. The remaining
piece, $J_1$, as usual, implements the Dirac equation on physical
fermionic states, and it is also conformally invariant.
We have constructed the leading corrections to $J$ and $J_1$ coming
from the $RR$ background. In principle one should expect corrections to
all orders in the $\alpha'$ expansion to each of these quantities.
As well, since there is no symmetry restricting the form of $Q$,
there may be additional terms of other ghost charges present. We suspect
that $Q$ in fact does truncate, as given, although we have not identified
the mechanism. 

We have not checked that $(Q_1)^2=0$ in the BRST cohomology of
$Q_0$, which would make the set of constraints consistent. We have found
that the conditions on massless vertex operators seem to be compatible
with supergravity results. This is an explicit check on the consistency
of the perturbation expansion.

Calculations become complicated at higher orders because we don't have
worldsheet supersymmetry as a guiding principle. Contact
 terms must be introduced order by order in the perturbation
expansion, and one would like a systematic approach in which to do
calculations, that is a prescription for the contact terms
to all orders. The guiding principle we have available is compatibility
of the $\alpha'$ expansion with spacetime supersymmetry; however, spacetime
supersymmetry  is corrected to all orders in $\alpha'$ as well, so in
practice it may be difficult to implement. Our \sigmod\ approach has given
us plausibility arguments that  there are no two loop corrections to the
$\beta$ function of the background in the $\alpha'$ expansion.
Given such a prescription, if one shows that one has compatibility 
with spacetime supersymmetry, one should be able to give a worldsheet 
proof of the non-renormalization theorems which have heretofore been 
based on spacetime supersymmetry arguments\cite{GrS, KR}.

The techniques in this paper if developed further should make it
possible to understand the spectrum of massive string states in  $RR$
backgrounds. Particularly, one should be able to compute the
perturbative spectrum of the string in the $AdS_5\times S^5$ geometry in
the infinite $N$ limit, as an expansion in the inverse of the t'Hooft
coupling, and it should depend only on defining the full OPE of the
fields appearing in the BRST current.

One should also be able to include the dilaton and the $NS$ $B$-field
for the most general supergravity background. The results should also
generalize without difficulty to any situation where we might have a
good knowledge of the CFT, as in the case of orbifolds.

Going to finite $N$ is more difficult, as it involves calculating the
one loop partition function of the string in these backgrounds.
Perturbation theory in $\alpha'$ is probably not sufficient for this
calculation, as one may need to go to all orders to understand issues
such as modular invariance, or a stringy exclusion principle\cite{MS}.

\acknowledgments We would like to thank the Aspen Center for
Physics and the Institute for Theoretical Physics at UCSB, where some of
this work was performed. We are indebted to Robert Dijkgraaf, Jacques Distler, David
Kutasov, Finn Larsen and Joe Polchinski for discussions of our results. RGL is supported
by an Outstanding Junior Investigator Award of the United States
Department of Energy, grant DE-FG02-91ER40677. Further partial support
was provided by NSF grant \# Phy94-07194.

\appendix

\section{Appendix: Conventions} {\label{Conventions}}
\renewcommand{\theequation}{A.\arabic{equation}}
\setcounter{equation}{0}

The fields in the superstring are described by the
spacetime coordinates $X^\mu$, their superpartners $\psi^\mu$, the
bosonic ghosts $b,c$ and the fermionic ghosts $\beta, \gamma$. One has
$\bar\psi^\mu,\bar b, \bar c, \bar\beta, \bar\gamma$ for the
antiholomorphic fields. One also includes the auxiliary field $F^{\mu}$
to close supersymmetry off shell.
The $\beta\gamma$ system is bosonized by
$\gamma= e^{\phi}\eta, 
\beta= e^{-\phi}\partial\xi$
with $\eta, \xi$ fermions of conformal dimension $1$ and $0$ respectively.

The OPE's of these fields that are used in
calculations around flat space are given as follows
\beqn
X^\mu(z) X^\nu(z') &=&-\eta^{\mu\nu} \log|z-z'|^2\\
\psi^\mu(z)\psi^\nu(z') &=&\eta^{\mu\nu}\frac 1{z-z'}\\ 
F^\mu(z) F^\nu(z') &=&
\eta^{\mu\nu}\partial\bar \partial \log|z-z'|^2 \\ S^\alpha
e^{-\phi/2}(z) S^\beta e^{-\phi/2}(z') &\sim&
{1\over z-z'}\Gamma^{\alpha\beta}_\mu\psi^\mu e^{-\phi} \\
\psi^\mu(z) S^\alpha(z') &\sim & \frac 1{(z-z')^{1/2}} 
\Gamma^{\mu\alpha}_{\dot\beta} S^{\dot\beta}\\
\psi^\mu(z) S^{\dot\alpha}(z') &\sim & \frac 1{(z-z')^{1/2}} 
\Gamma^{\mu\dot\alpha}_{\beta} S^{\beta}
\eeqn
with similar OPE's for the right moving fields.

Divergent quantities are defined through
\begin{equation} 
\langle X(z)X(z)\rangle = \log\epsilon =
\int \frac{d^2p}{(2\pi)^2 |p|^2} 
\end{equation} 
It is easy to show also that 
\begin{equation} 
\int
\frac{d^2z}{|z-z'|^2} = \log{\epsilon} 
\end{equation}

For curved spaces, fermions are bosonized by pairing in an orthonormal
frame 
\begin{equation} 
\psi^\mu = e^\mu_a \chi^a 
\end{equation} 
and it
is $\chi^a$ that is bosonized in a standard fashion. With this
bosonization it is possible to define spin fields in the curved
manifold (with respect to the given orthonormal frame).
The OPE of these spin fields is 
\begin{equation} 
S^\alpha
e^{-\phi/2}(z)S^\beta e^{-\phi/2}(z') \sim \frac{\Gamma_a^{\alpha\beta}
}{z-z'} \chi^a e^{-\phi} 
\end{equation} 

The picture changing operator is defined by
\beq
P_{+1}= Q_{BRST} \xi
\eeq

In flat backgrounds, the relevant piece of the picture changing operator
for integrated vertex operators is given by
\beq
P_{+1}\sim e^\phi
\psi^\mu\partial X^\mu = e^\phi G_m
\eeq
We also formally define $\phf$ to be the holomorphic square root of the
picture changing operator
\beq
\phf^2= P_{+1}
\eeq
The relevant piece of the picture-changing operator in a curved geometry
will be 
\begin{equation} 
P_{+1}\sim e^\phi\chi^a\pa\xi^a\sim e^\phi \psi^\nu\partial X^\mu g_{\mu\nu}(X),
\end{equation}
where $\xi^a$ is the normalized tangent vector. This 
result which is independent of the $RR$ background to first
order in the $\alpha'$ loop expansion. 

Occasionally, we also use the shorthand notation
\begin{equation}
\CS^{\alpha} = P_{1/2} e^{-\phi/2}S^\alpha
\end{equation}
with a similar expression for the right moving fields
$\bar\CS^\beta$.

\medskip

%\bibliography{BRST} \bibliographystyle{uiuchept}

\providecommand{\href}[2]{#2}\begingroup\raggedright\endgroup

\end{document}